# AN INTEGRATED CIRCUIT FOR SIGNAL PROCESSING OF THE AMS RICH PHOTOMULTIPLIER TUBES

A. Barrau, L. Gallin-Martel, J. Pouxe, O. Rossetto
Institut des Sciences Nucléaires CNRS-IN2P3 & Université Joseph Fourier
53 Avenue des Martyrs 38 026 Grenoble-Cedex France

*Abstract*

An analog integrated circuit has been designed, in a BiCMOS 0.8 micron technology, for the feasability study of the signal processing of the AMS RICH photomultiplier tubes [1]. This low power, three channel gated integrator includes its own gate generator and no external analog delay is required. It processes PMT pulses over a dynamic range of 100. An external RC network allows it to work with various PMT gains. It provides a logic output that indicates whether the associated analog output has to be considered. This gated integrator is used with a compact DSP based acquisition system in a 132 channel RICH prototype. The charge calibration of each channel of the RICH is carried out using the light of a LED. The pedestal measurement is performed on activation of a dedicated input. The noise contribution study of the input RC network and of the amplifier's is presented.

## I. INTRODUCTION

Although the RICH for the AMS project is not today completely defined, it is reasonable to think that it will include between 3000 and 10000 PMTs. This large number of detector cells and the low power consumption required by the spacial constraints lead to the integration of the analog processing part of the PMT signals. The simulation results [2] show that the number of photons detected by each PMT can reach 50 for heavy ions beyond oxygen in a 1x1 $cm^2$ pixel. Thus the electronic dynamic range must be from 1 to about 100. At least two methods can deliver information on the number of photons detected by a PMT. The first one based on the time over threshold measurement, after a pulse stretching, had been proposed for the AMS RICH [3]. The main drawback of this method is that a small offset on the voltage comparator input produces a large error on the final measurement. The offset has to be carefully adjusted and controlled. As the number of channels is quite large, this adjustment can become onerous. The other method which we retained is the PMT pulse charge integration. This self gated integrator includes its own gate generator and no analog delay is required. An external RC network performs the stretching of the PMT pulse and minimizes the non integrated part of the input signal. The error introduced by this processing method depends mainly on the comparator speed. The worst case is reached for the single photon detection but even for this case the error is negligible. The block diagram of one channel of the integrated circuit is given in Figure 1. Some circuits performing a charge to voltage conversion have already been designed, but they require an external gate signal and their power consumption is not compatible with the AMS constraints [4].

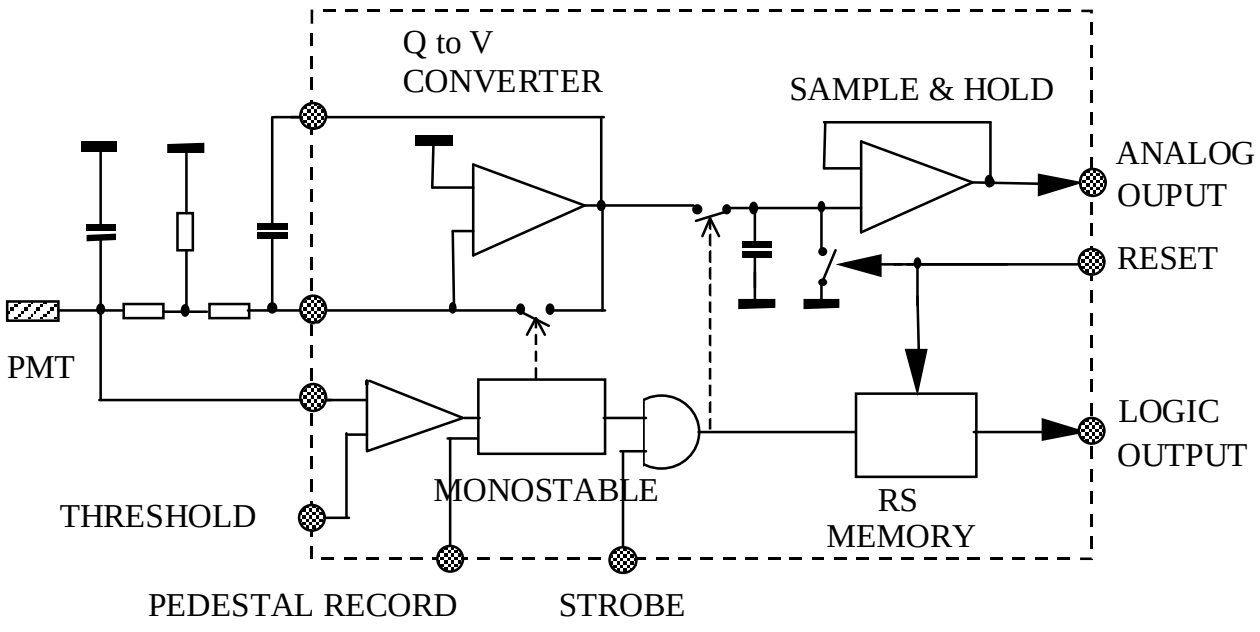

Figure1 : Block diagram of the PMT pulse processing

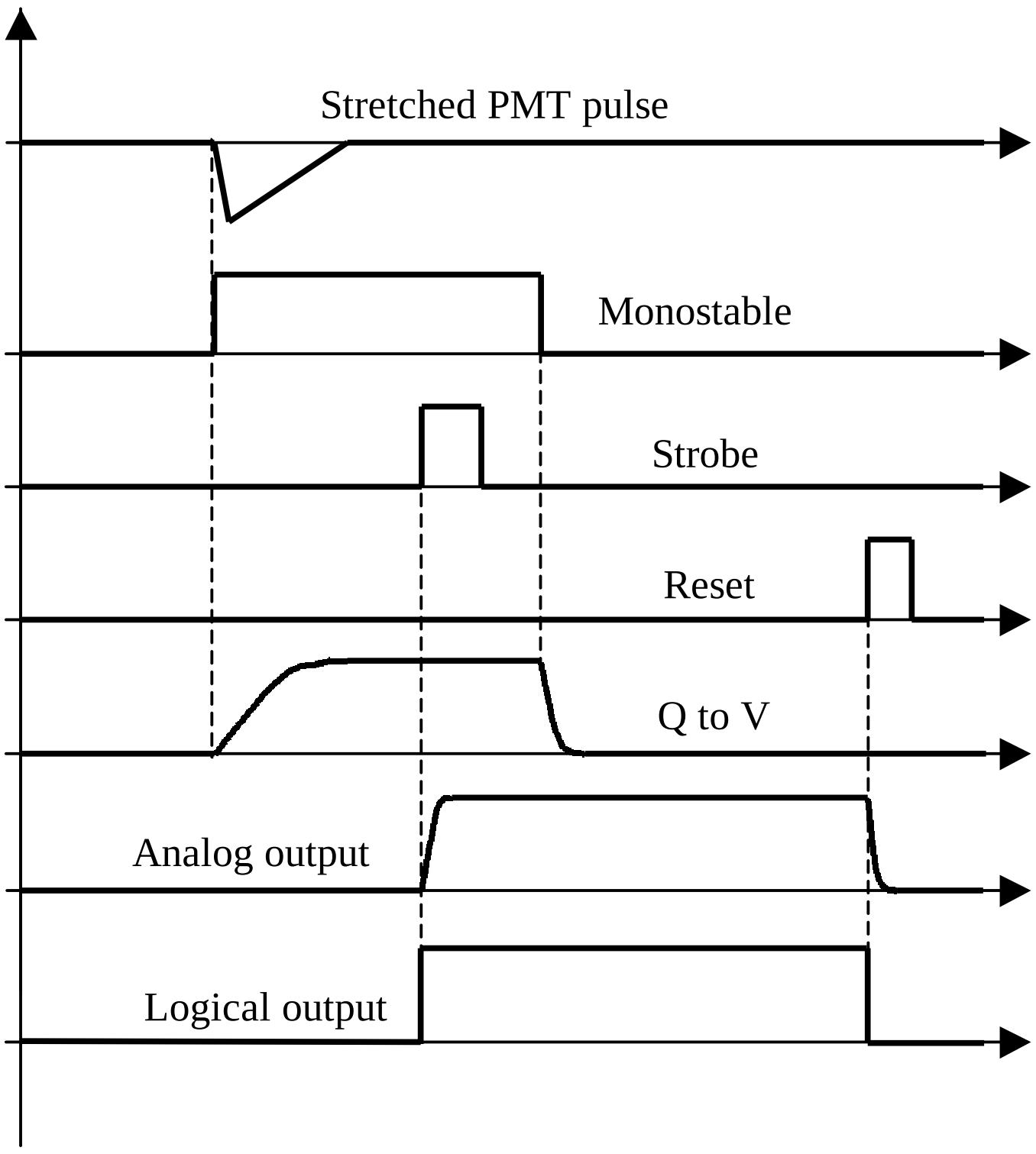

Figure 2 : Circuit chronograms

This integrated circuit is composed of five parts :

high speed voltage comparator

monostable

charge to voltage converter

sample and hold

register

When a pulse occurs at the output of a PMT (figure 2), the voltage comparator triggers on the monostable that enables the charge to voltage converter to integrate the input charge during a period fixed by the monostable. The dynamic range of the integrated charge is from 0.2 pC for the single photoelectron to 20 pC. If the STROBE signal occurs during the monostable pulse, the output of the charge to voltage converter is transferred to the sample and hold circuit. This coincidence also sets a RS flipflop which indicates that the analog output has to be stored after digitalization. The switches shown in Figure 1 are built with two complementary MOS transistors to reduce the ON resistance and the injected charges. The sample and hold circuit stores the result of the integration until it is converted by an external ADC. The converter (AD9220 a 12-bit, 10 MSPS, 250mW ADC) uses a pipelined architecture and processes 32 analog channels. It is located on the same printed board as the ASICs. The STROBE signal provided by the RICH trigger enables the conversion of each analog channel and the storage of the data depends on the state of the logical output [5].

## II. VOLTAGE COMPARATOR

The comparator must be of high speed and high gain in order to enable the integrator with minimum delay to minimize the non integrated part of the stretched PMT pulse.

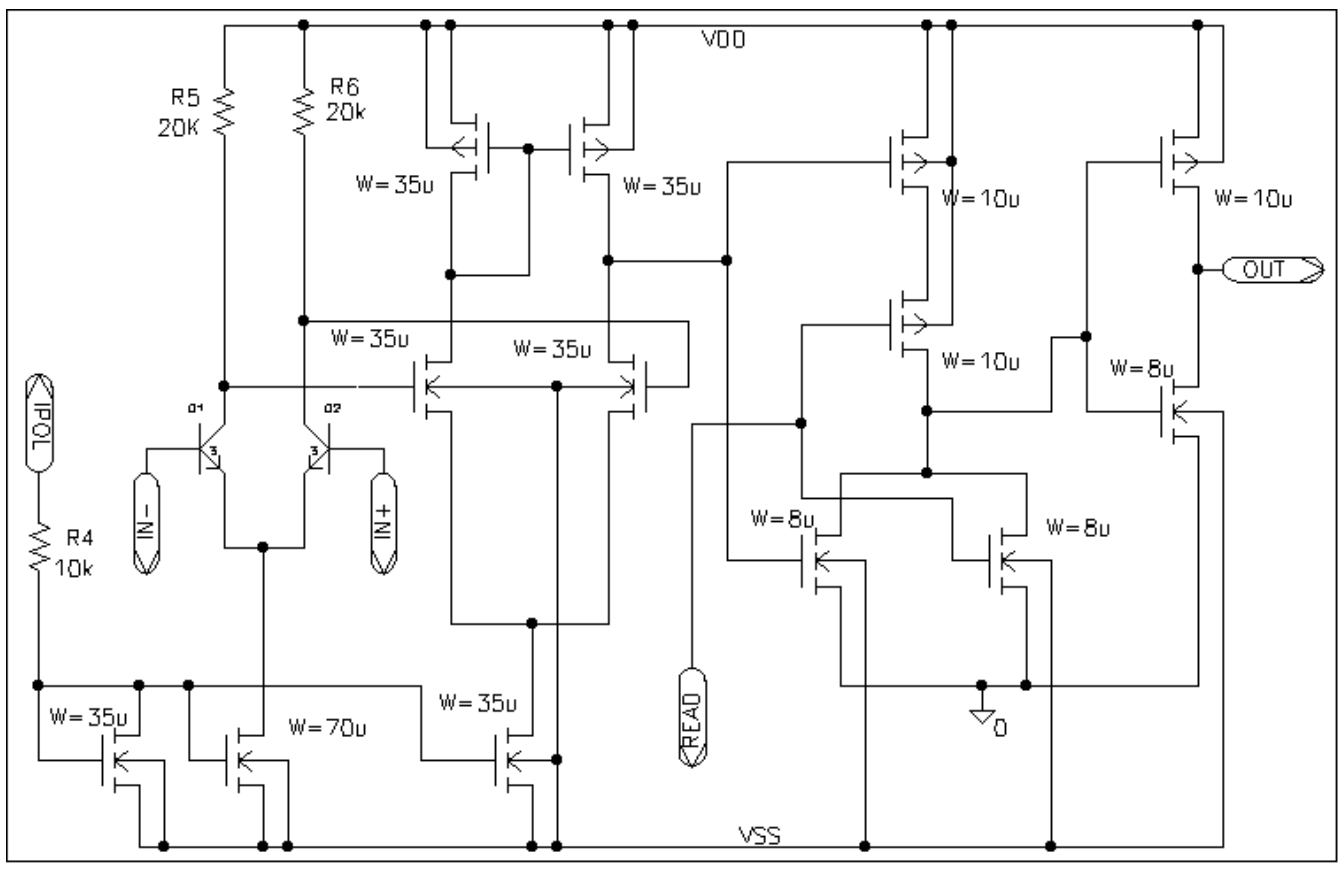

Figure 3 : Comparator schematics

A passive network performs the shaping of the PMT pulse (Figure 1). The pulse amplitude depends mainly on the value of the capacitor. This capacitance of about 15 pF stands for the 10 cm coaxial cable and the printed circuit board capacitance. The comparator must have a dynamical range of 100 and the pulse amplitude, limited by the negative power supply, must be lower than -3V. The single photon amplitude has to be set at a value of -30mV to meet these conditions. The PMT gain can be directly deduced from the expression :

$$G \approx \frac{C \cdot V}{1.6 \cdot 10^{-19}} = \frac{15 \cdot 10^{-12} \cdot 30 \cdot 10^{-3}}{1.6 \cdot 10^{-19}} \approx 3 \cdot 10^{6}$$

The threshold is set at -10 mV which corresponds to one third of the single photon mean amplitude. The inverting and non inverting inputs (Figure 3) are accessible allowing the integrator to be enabled by an anode or a dynode pulse. This provides the possibility to integrate larger pulses without enabling the comparator above the power supply. The typical delay is 3.5 ns and the time walk less than 4 ns over the whole dynamical range (Figure 4). The time walk is a critical point for the linearity in small signals.

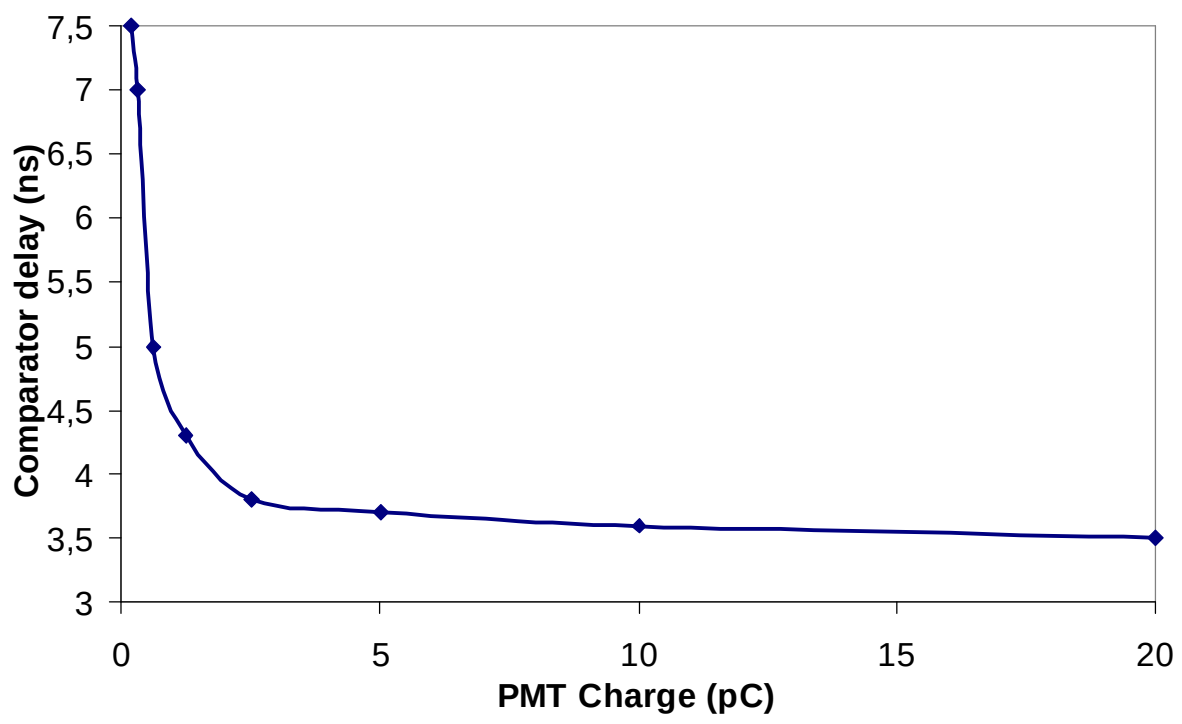

Figure 4 : Measured comparator delay

The comparator can also be triggered on by the READ input (Figure 3) which allows the measurement of the analog output pedestal.

## III. CHARGE TO VOLTAGE CONVERTER

The gated integrator, or charge to voltage converter, is built using a charge preamplifier structure. In the absence of an enable signal from the monostable, the MOS switch is on, so the loop gain is about –1. When a pulse occurs, the monostable opens the MOS switch which allows the integration of the PMT charge by the amplifier and its feedback capacitor. The resistive part of the input network performs a current division in order to avoid the saturation of the integrator. Only one half of the PMT charge is integrated**.** The feedback capacitor is the critical component that gives the accuracy of the analog processing. As integrated structures have generally a poor precision (typically ± 15%), this capacitor is external to the chip. This configuration allows the adjustment of the integrator gain to various PMT gains. The two gain stage structure provides an open loop gain of about 2500. The effect of the open loop gain of the amplifier on the noise and the feedback capacitor discharge has been studied and is summed up below.

### *A. Noise analysis*

The circuit shown in figure 5 was used to study the effect of the amplifier open loop gain. This gain is supposed to be of the first order :

$$A = \frac{A_0}{1 + \tau p}$$

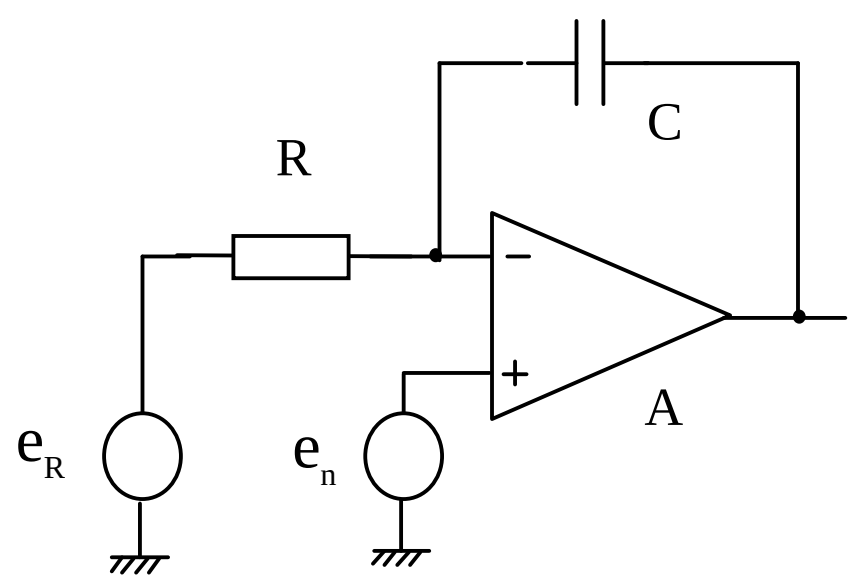

Figure 5 : Noise analysis schematics

The equivalent output noise can be calculated by summing the contribution of the resistor and that of the amplifier itself.

*A1. Amplifier white noise contribution*

For a voltage source the transfer function is :

$$\frac{V_n}{e_n} = A\,\frac{RCp + 1}{(A+1)RCp + 1}$$

$$\frac{V_n}{e_n} = A_0\,\frac{RCp+1}{RC\tau p^2 + p[\tau + (A_0+1)RC] + 1}$$

$$\Delta = [(A_0+1)RC + \tau]^2 - 4RC\tau > 0$$

since $\Delta > 0$ the previous expression can be written as

$$\frac{V_n}{e_n} = \frac{A_0}{\tau}\cdot\frac{p+a}{(p+b)(p+c)}$$

with:
$$\begin{cases} a = \dfrac{1}{RC} \\ b = \dfrac{(A_0+1)RC + \tau + \sqrt{\Delta}}{2RC\tau} \\ c = \dfrac{(A_0+1)RC + \tau + \sqrt{\Delta}}{2RC\tau} \end{cases}$$

The noise spectral density is :

$$V_n^2 = \frac{A_0^2}{\tau^2}\cdot\frac{\omega^2 + a^2}{(\omega^2+b^2)(\omega^2+c^2)}e_n^2$$

with $e_n^2 = \dfrac{\alpha}{2\pi}d\omega$

And the square of the rms noise is :

$$V_{nrms}^2 = \frac{A_0^2}{\tau^2}\cdot\frac{\alpha}{2\pi}\cdot\int_0^\infty \frac{\omega^2 + a^2}{(\omega^2+b^2)(\omega^2+c^2)}d\omega$$

$$V_{nrms}^2 = \frac{A_0^2}{\tau^2}\cdot\frac{\alpha}{4}\cdot\frac{a^2+bc}{(b+c)bc}$$

$$V_{nrms}^2 = \frac{A_0^2}{\tau}\cdot\frac{\alpha}{4}\cdot\frac{\tau + RC}{(A_0+1)RC + \tau}$$

assuming $\tau >> RC$ and $\tau \approx \beta A_0$

$$V_{nrms}^2 \approx \frac{\alpha}{4}\cdot\frac{A_0}{RC+\beta}$$

*A2. Amplifier Flicker noise*

The input of charge to voltage converter is AC coupled to the external RC network. This 100 nF serial capacitor is not shown on Figure 1. The purpose of this capacitor is :

To avoid a leakage current due to the amplifier offset and due also to the comparator biasing input current.

To limit the contribution of the amplifier Flicker noise.

*A3. Resistor noise contribution*

The noise spectral density is :

$$V_R^2 = \frac{A_0^2}{R^2C^2\tau^2}\cdot\frac{1}{(\omega^2+b^2)(\omega^2+c^2)}e_R^2$$

with $e_R^2 = \dfrac{4k\,\mathrm{TR}}{2\pi}d\omega$

$$V_{Rrms}^2 = \frac{A_0^2}{R^2C^2\tau^2}\cdot\frac{4k\,\mathrm{TR}}{2\pi}\cdot\int_0^\infty\frac{1}{(\omega^2+b^2)(\omega^2+c^2)}d\omega$$

$$V_{Rrms}^2 = \frac{A_0^2 k\,\mathrm{TR}}{R^2C^2\tau^2}\cdot\frac{1}{bc(b+c)}$$

$$V_{Rrms}^2 = \frac{A_0^2 k\,\mathrm{TR}}{(A_0+1)RC+\tau}$$

$$V_{Rrms}^2 = \frac{A_0 k\,\mathrm{TR}}{RC+\beta}$$

$$V_{rms}^2 = V_{nrms}^2 + V_{Rrms}^2 = \frac{A_0}{RC+\beta}\left(\frac{\alpha}{4} + k\,\mathrm{TR}\right)$$

The total rms noise is :

$$V_{rms} = \sqrt{\frac{A_0}{RC+\beta}\left(\frac{\alpha}{4} + k\,\mathrm{TR}\right)} \qquad (1)$$

The rms noise depends on the square root of the open loop gain.

The simulated value of the amplifier white noise is :

$$e_n = 4.5\,nV/\sqrt{Hz}$$

For a R =1 kΩ and C =10 pF, the total rms noise is :

$V_{rms} \approx 1mV$ with $Q_{rms} = CV_{rms} = 0.01pC$

This value corresponds to 0.05 photoelectron.

## *B. Feedback capacitor discharge analysis*

The circuit shown in Figure 6 was used to study the capacitor discharge. In order to simplify the calculation the signal of PMT is supposed to be a $t_0$ duration square pulse.

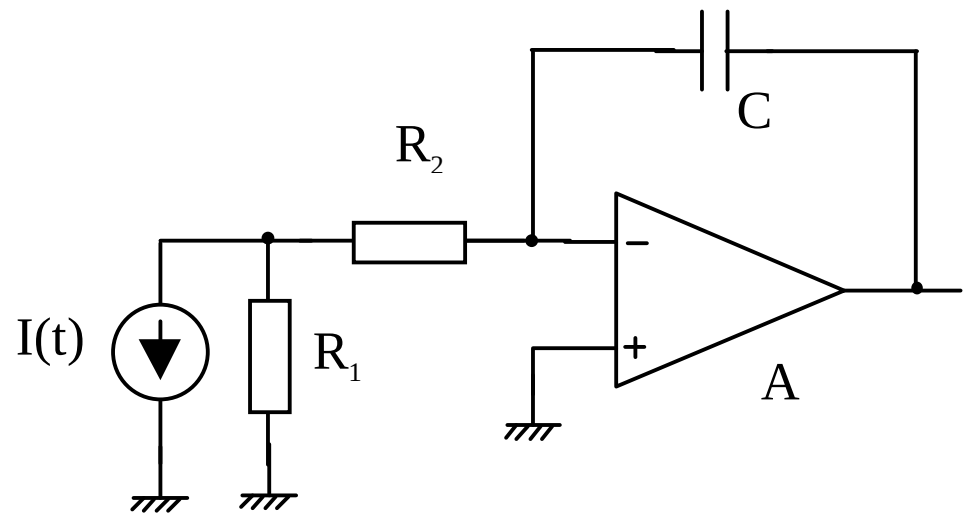

Figure 6 : Schematic of the capacitor discharge analysis

In this case the open loop gain of the amplifier is the DC gain so $A = A_0$. The output voltage as a function of the input current is:

$$V_s = \frac{-A_0 i(t) R_1}{(A_0+1)(R_1+R_2)Cp+1}$$

The Laplace transform of the input current is given by:

$$I(p) = \frac{I_0}{p}(1-e^{-pt_0})$$

This leads to:

$$V_s = A_0 R_1 I_0 (e^{-t/\tau} - e^{-(t-t_0)/\tau})$$

with $\tau = (A_0+1)(R_1+R_2)C$ and $t \geq t_0$

Assuming t and $t_0$ are small compared to $\tau$ an approximation of the previous expression is :

$$Vs = A_0 R_1 I_0 \left(\frac{-t_0}{\tau} + \frac{tt_0}{\tau^2} - \frac{t_0^2}{2\tau^2}\right)$$

The discharge slope of the capacitor is given by :

$$\frac{dVs}{dt} = \frac{A_0 R_1 I_0 t_0}{\tau^2} \approx \frac{R_1 I_0 t_0}{A_0 (R_1+R_2)^2 C^2} \qquad (2)$$

The discharge slope depends on the inverse of the DC open loop gain. Expressions (1) and (2) show the necessity of choosing a compromise value for the amplifier open loop gain.

## III. MONOSTABLE

The monostable provides a pulse with a fixed duration from the output of the comparator (Figure 7). This signal directly controls the integration duration. The leading edge of the comparator output is differentiated by the capacitor C1 and used to charge the capacitor C2 through the MOS transistor M1. Then this capacitor is discharged under a constant current. The capacitor voltage is compared to a reference voltage then, when the discharge reaches this voltage the output of the monostable returns to zero.

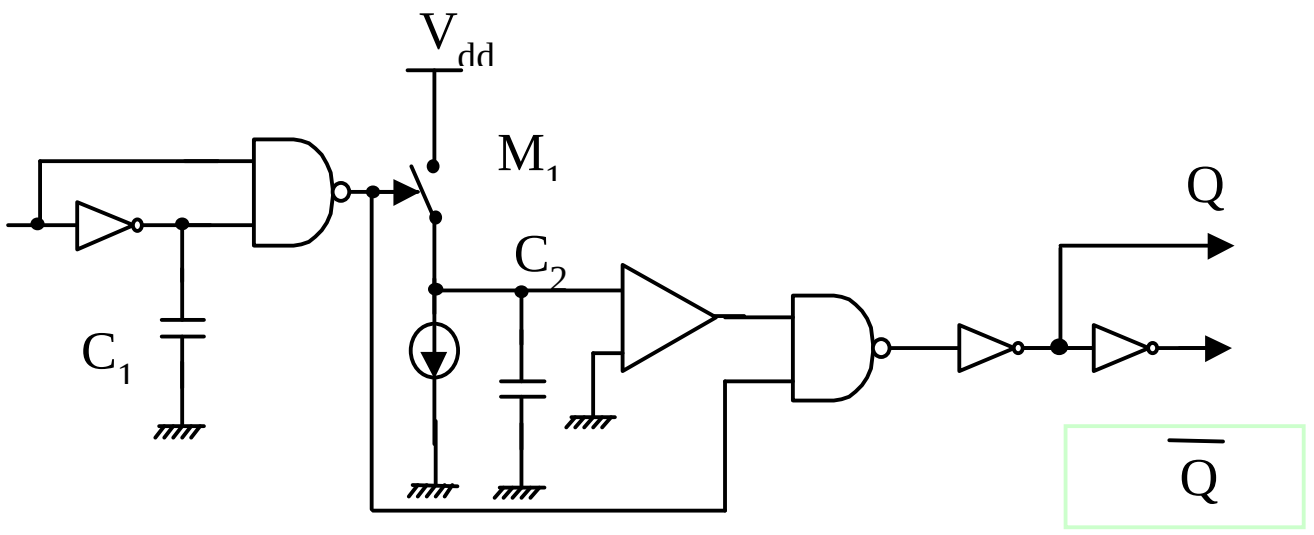

Figure 7: Monostable principle

The monostable duration is fixed by the capacitor discharge current. This one can be adjusted by a single external resistor for the three channels. In order to minimize the delay of the charge to voltage converter the leading edge of the monostable output is directly driven by the differentiated comparator output.

## IV. SAMPLE AND HOLD

The sample and hold circuit stores the voltage value of the integrator until this value is transfered to the ADC. The amplifier of the sample and hold circuit has a MOS transistor input in order to reduce the leakage current of the hold capacitor. The storing time required for the sample and hold circuit is determined by the speed of the ADC. With a 3pF hold capacitor, a storing time larger than 100 µs is achieved without significant voltage loss at the output. The ADC is multiplexed with 32 analog channels. Its pipelined architecture leads to convert each channel and the digital outputs of the integrator indicate which analog output has to be stored after digitalization.

## V. MAIN FEATURES OF THE CIRCUIT

The main charateristics of the circuit are :

dynamical range : 100 photoelectrons

integral non linearity : < ± 0.5%

sample and hold leakage : < ± 5 µV/µs

self gated integrator

external integration capacitor

power consumption : 6 mW / channel

power supplies : +3.3V and –2V

0.8 micron BiCMOS technology

PLCC28 package

The integral non linearity was measured using a pulse generator, providing a pulse similar to the stretched PMT pulse, and a precision attenuator. Figure 8 shows the INL over a dynamical range of 100 (from 0.2 pC to 20 pC).

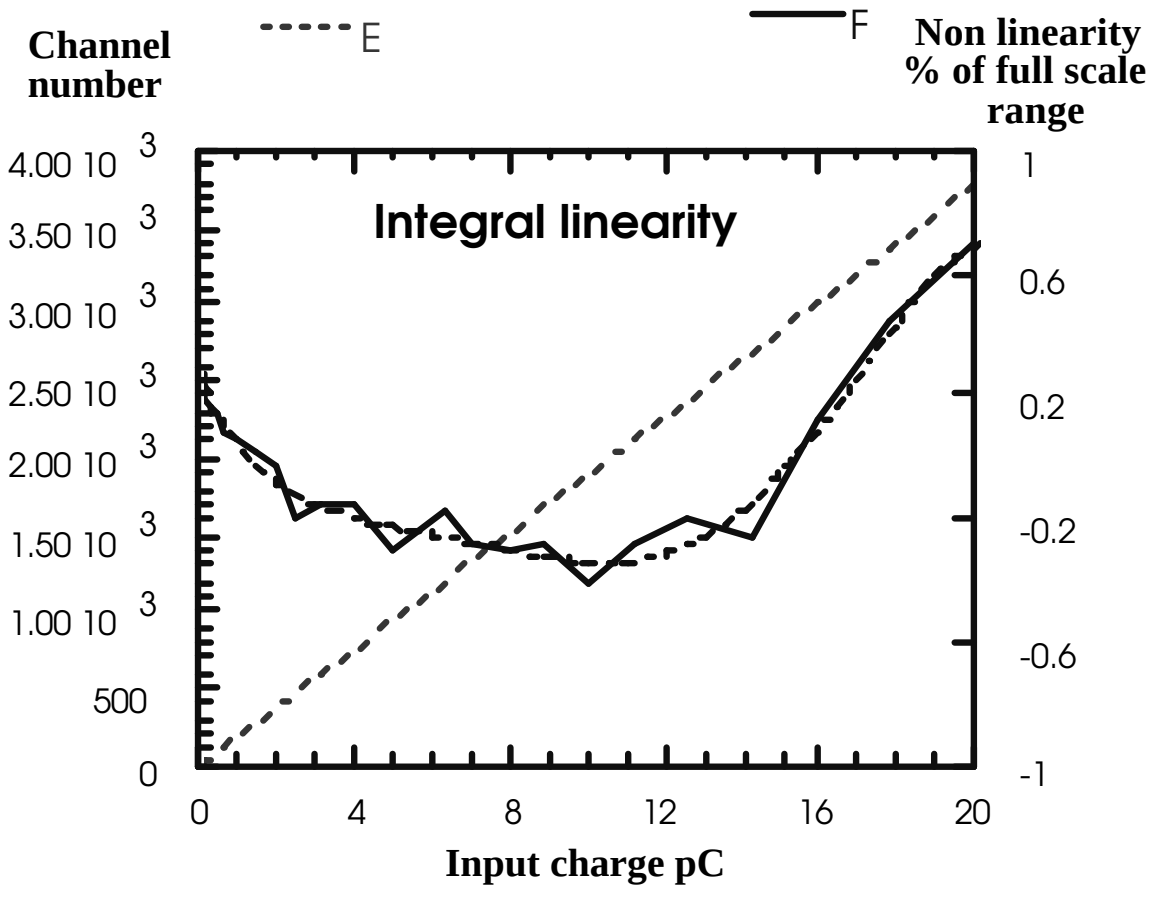

Figure 8 : Integral linearity over a dynamical range of 100

The charge calibration of each channel may be carried out using the light of a LED. If the PMT has a good single photon response it is easy to locate the position of the first and the second photoelectrons. Figure 9 shows the spectrum and the Gaussian fits of a XP2802 PMT response.

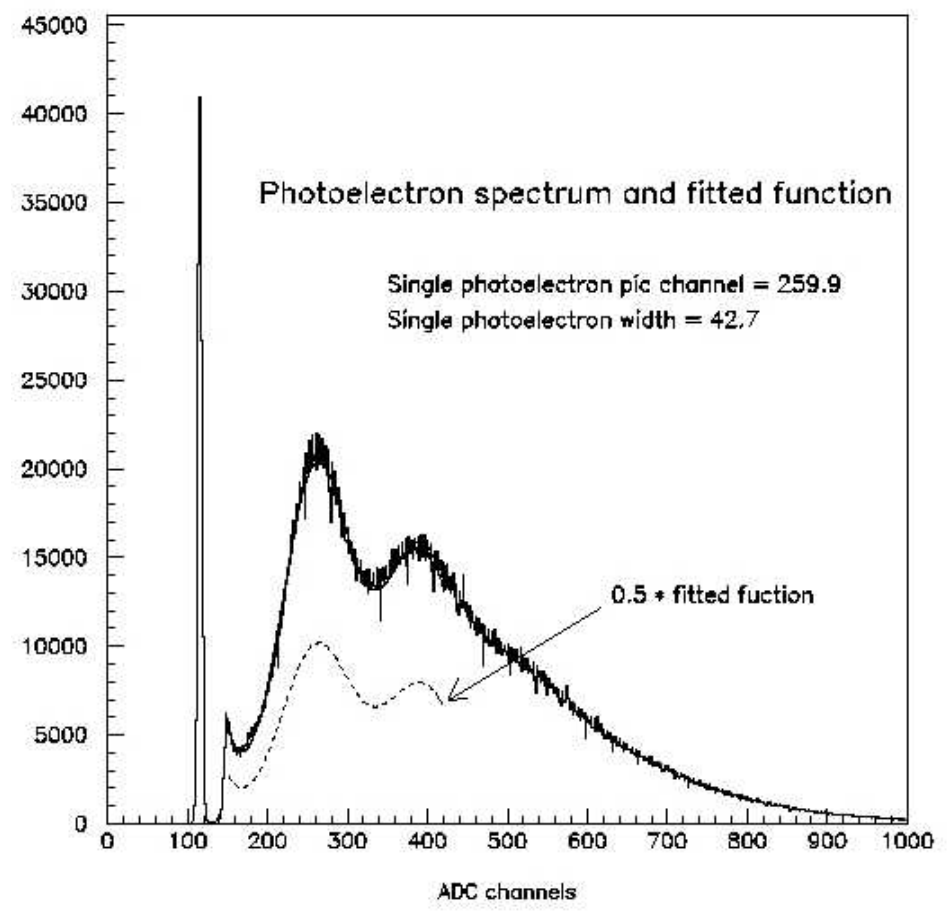

Figure 9 : Pedestal and photon distributions

The fitted function can be written as :

$$f(x) = K_1 \sum_{i=1}^{N} Q_i e^{-\frac{(x-iP+(i-1)ped)^2}{2i\sigma^2}} + K_2 e^{-J(x-ped)}$$

In practice the infinite sum is limited to the 3 first terms (N=3). $K_1$ and $K_2$ are just normalization factors, $Q_i$ is the respective amplitude of the $i^{th}$ photoelectron. It can't be statistically imposed by a Poisson statistics because of the intrinsic LED power variations. On the other hand the width of the $i^{th}$ Gaussian is entirely determinated as : $\sigma\sqrt{i}$ .

The exponential term represents the discrete processes (thermoemission, noise, ...). Among the N+5 free parameters ($K_1$, $K_2$, P, $\sigma$, J, $Q_i$), the two physical ones are P, the single photoelectron pic position and $\sigma$, the single photoelectron pic width. It should be pointed out that such a method is not very sensitive to the input parameters.

## VI. Layout

The circuit has been integrated using the Austria Mikro Systeme (AMS) 0.8 micron BiCMOS technology. The chip integrates three identical channels on an 2x2 $mm^2$ area

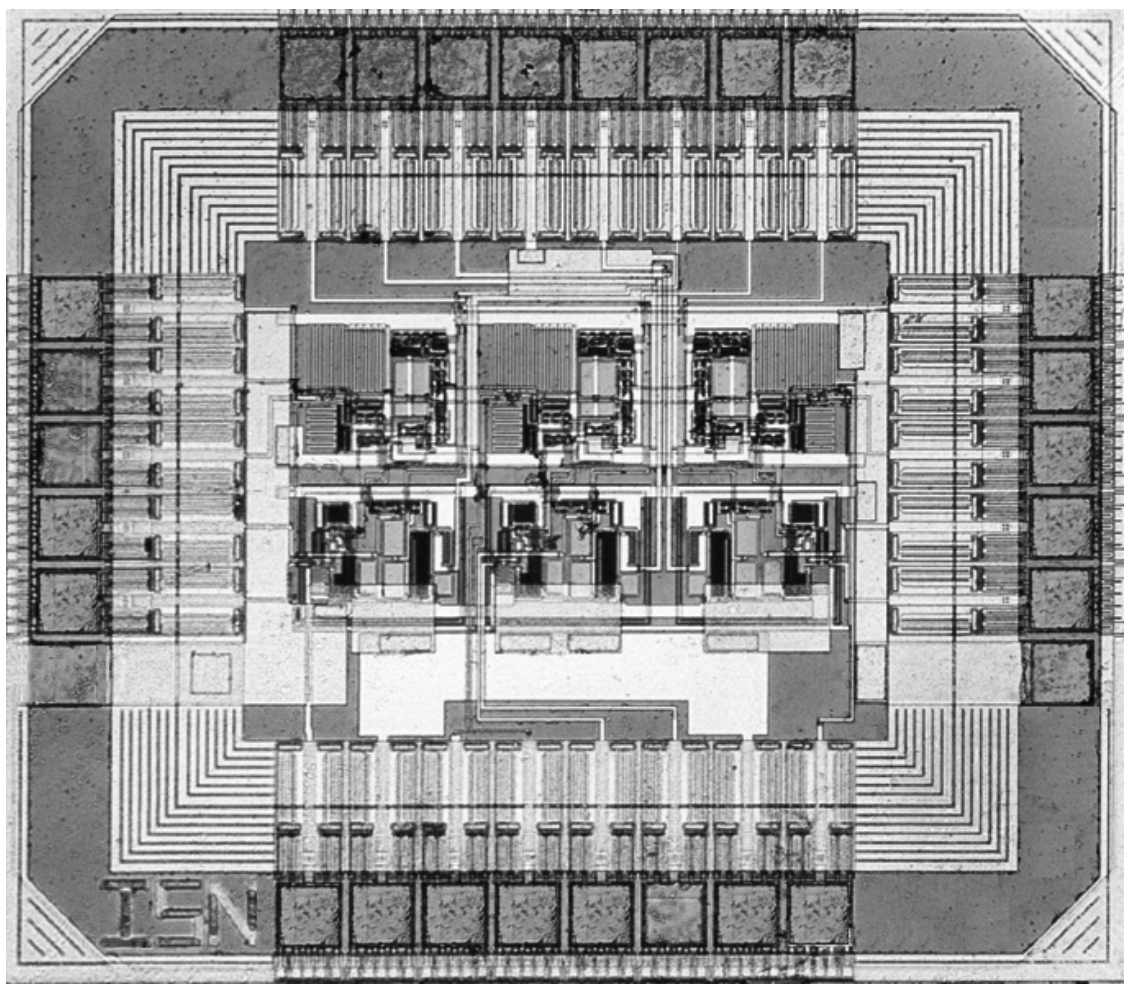

Figure 10 : Layout

## VII. Conclusion

This integrated circuit, with its low power consumption and its good linearity presents characteristics compatible with the AMS RICH project. The circuit is already in use, with the DSP based acquisition system, on a reduced scale prototype of the RICH (132 PMT channels) on which the whole system is going to be validated.